\documentclass[global,twocolumn]{svjour}
\usepackage{graphicx,color,xspace,multirow,amssymb}
\usepackage{textcomp}
\usepackage{amsmath}
\usepackage{fixltx2e}
\journalname{Applied Physics B}

\def\fm#1{\ifmmode #1 \else $#1$\fi}
\def\ket#1{{%
  \ifmmode |\,#1\,\rangle \else $|\,#1\,\rangle$\fi}}
\def\bra#1{{%
  \ifmmode \langle\,#1\,| \else $\langle\,#1\,|$\fi}}

\def\nCa40{\fm{^{40}\mathrm{Ca}}\xspace}
\def\Ca40{\fm{\nCa40^{+}}\xspace}

\def\ms{\fm{{}\mathrm{S}}\xspace} 
\def\mP{\fm{{}\mathrm{P}}\xspace} 
\def\md{\fm{{}\mathrm{D}}\xspace} 
\def\ssoh{\fm{{}\mathrm{S}_{1/2}}\xspace} 
\def\spoh{\fm{{}\mathrm{P}_{1/2}}\xspace}
\def\sdth{\fm{{}\mathrm{D}_{3/2}}\xspace}
\def\spth{\fm{{}\mathrm{P}_{3/2}}\xspace}
\def\sdfh{\fm{{}\mathrm{D}_{5/2}}\xspace}

\def\sohmmoh{\fm{{}\ket{\mathrm{S}_{1/2}, - \frac{1}{2}}}\xspace} 
\def\sohmpoh{\fm{{}\ket{\mathrm{S}_{1/2}, + \frac{1}{2}}}\xspace}

\def\pthmmth{\fm{{}\ket{\mathrm{P}_{3/2}, - \frac{3}{2}}}\xspace}
\def\pthmmoh{\fm{{}\ket{\mathrm{P}_{3/2}, - \frac{1}{2}}}\xspace}

\def\dfhmmfh{\fm{{}\ket{\mathrm{D}_{5/2}, - \frac{5}{2}}}\xspace}
\def\dfhmmth{\fm{{}\ket{\mathrm{D}_{5/2}, - \frac{3}{2}}}\xspace}

\def\dfhmpoh{\fm{{}\ket{\mathrm{D}_{5/2}, + \frac{1}{2}}}\xspace}
\def\dfhmpth{\fm{{}\ket{\mathrm{D}_{5/2}, + \frac{3}{2}}}\xspace}
\def\dfhmpfh{\fm{{}\ket{\mathrm{D}_{5/2}, + \frac{5}{2}}}\xspace}

\def\dfhmpmth{\fm{{}\ket{\mathrm{D}_{5/2}, \pm \frac{3}{2}}}\xspace}
\def\dfhmpmfh{\fm{{}\ket{\mathrm{D}_{5/2}, \pm \frac{5}{2}}}\xspace}
\def\mstmp{\ms\fm{\leftrightarrow}\xspace\mP} 

\def\mptmd{\mP\fm{\leftrightarrow}\xspace\md}
\def\mstmd{\ms\fm{\leftrightarrow}\xspace\md}
\def\ssohtspoh{\ssoh\fm{\leftrightarrow}\xspace\spoh} 
\def\ssohtspth{\ssoh\fm{\leftrightarrow}\xspace\spth}
\def\spohtsdth{\spoh\fm{\leftrightarrow}\xspace\sdth}

\def\spthtsdfh{\spth\fm{\leftrightarrow}\xspace\sdfh}

\def\sdfhtspth{\sdfh\fm{\leftrightarrow}\xspace\spth}
\def\ssohtsdfh{\ssoh\fm{\leftrightarrow}\xspace\sdfh}
\def\ssohtsdth{\ssoh\fm{\leftrightarrow}\xspace\sdth}

\def\ramantrans{\ssoh\fm{\leftrightarrow}\xspace\spth\fm{\leftrightarrow}\xspace\sdfh} 

\def\mus{\fm{\mu\mathrm{s}}\xspace}
\def\mum{\fm{\mu\mathrm{m}}\xspace}

\begin{document}
\title{Toward an ion-photon quantum interface in an optical cavity}
\author{a lot of names}
\author{A.~Stute\inst{1} \and B.~Casabone\inst{1} \and B.~Brandst\"{a}tter\inst{1} \and D.~Habicher\inst{1} 
\and H.~G.~Barros\inst{1}\thanks{\emph{Present address:} Department of Physics and Center  for Applied Photonics, Universit\"at Konstanz, 78457 Konstanz, Germany}
\and P.O.~Schmidt\inst{1}\thanks{\emph{Present address:} QUEST Institute for Experimental Quantum Metrology, PTB and Leibniz University of Hannover, 38116 Braunschweig, Germany} \and T.E.~Northup\inst{1} \and  R.~Blatt\inst{1,2}
}                     
\institute{Institut f\"{u}r Experimentalphysik, Universit\"{a}t Innsbruck, 6020 Innsbruck, Austria \and Institut f\"{u}r Quantenoptik und Quanteninformation (IQOQI), 6020 Innsbruck, Austria}
\mail{tracy.northup@uibk.ac.at}

\date{Received: date / Revised version: date}
%
\maketitle
\begin{abstract}
We demonstrate several building blocks for an ion-photon interface based on a trapped \Ca40 ion in an optical cavity.  We identify a favorable experimental configuration and measure system parameters, including relative motion of the trapped ion and the resonator mode.  A complete spectrum of cavity-assisted Raman transitions between the  $4^{2}\ssoh$ and $3^{2}\sdfh$ manifolds is obtained.  On two of these transitions, we generate orthogonally polarized cavity photons, and we demonstrate coherent manipulation of the corresponding pair of atomic states.  Possible implementations of atom-photon entanglement and state mapping within the ion-cavity system are discussed. 
\end{abstract}

\section{Introduction} \label{introduction}

Trapped, laser-cooled ions are excellent candidates for quantum-information processing \cite{Blatt08}, and ion systems have been shown to fulfill all five DiVincenzo criteria for quantum computation \cite{DiVincenzo00,ARDA04}.  Two additional DiVincenzo criteria stipulate what is necessary for quantum communication, that is, for a quantum network linking quantum computers  \cite{DiVincenzo00}.   These criteria envision a deterministic network, based on a coherent interface between ``stationary'' and ``flying'' qubits as well as transmission of flying qubits between remote locations.  Possible ion-based realizations of a quantum network are diverse, including microtraps in which ions are steered between storage and interaction regions \cite{Kielpinski02} and wires in which motion-induced currents couple two spatially separated ions \cite{Daniilidis09}.   A quantum network may also be probabilistic; in this case, it is the process of measurement that accomplishes long-distance transmission of qubits \cite{Duan10}.  Key ingredients for a probabilistic ion-trap quantum network have recently been demonstrated, including both entanglement and a heralded gate between remote ions \cite{Moehring07,Maunz09}.  Here, we focus on the deterministic scheme from the original proposal for ion-based quantum networks \cite{Cirac97}, in which photons act as flying qubits that transfer quantum information between stationary ions.  

In order for information transfer to be reversible in the scheme of Ref. \cite{Cirac97}, individual ions are placed inside separate optical resonators, and a quantum interface is realized via the coherent coupling of the ion to the resonator field.  In practice, achieving a strong ion-field coupling is quite challenging due to the difficulty of integrating high-finesse mirrors within an ion-trap apparatus.  While the strongest coupling is achieved with small-mode-volume cavities, where mode volume is inversely proportional to cavity length, ion-trap cavities to date have only been constructed on the cm scale \cite{Guthoehrlein01,Mundt02,Russo09,Leibrandt09,Herskind09}. In contrast, cavities of just tens of $\mu$m in length are possible in experiments with neutral atoms \cite{Miller05}.  Thus, it is in the context of neutral atoms that the greatest progress toward a cavity-based interface has been demonstrated, including the generation of sequential polarization-entangled single photons \cite{Wilk07b} and the coherent mapping of a photonic state onto an intracavity atom \cite{Boozer07a}.  Nevertheless, trapped ions offer several advantages as stationary qubits, including long storage times and the fact that they can be coherently manipulated and detected with high fidelity.  Within a cavity, ion state manipulation and readout enable the preparation of target ion-cavity states and analysis not only of cavity photons but also of the atomic component.  In addition, a single ion can be positioned precisely with respect to the cavity mode and localized to $\sim10$ nm via ground-state cooling in three dimensions.

We propose an ion-photon interface within an optical cavity using \Ca40, based on the parameters of the experiment described in Ref. \cite{Russo09}.  After selecting advantageous atomic and photonic states for this interface, we present results that lay the groundwork for it, including spectroscopy of the $4S_{1/2}-4P_{3/2}-3D_{5/2}$ cavity-assisted Raman transition, orthogonally polarized photons generated on two target transitions, and coherent manipulation of the atomic qubit.  We outline the necessary steps to realize atom-photon entanglement in this system and consider viable further applications.

\section{Theoretical and technical considerations} \label{transitions}
  \subsection{States and transitions for an atom-photon interface in \Ca40} \label{transitions_in_ca}
\begin{figure}
\centering
  \resizebox{0.4\textwidth}{!}{%
    \includegraphics[width=0.4\textwidth]{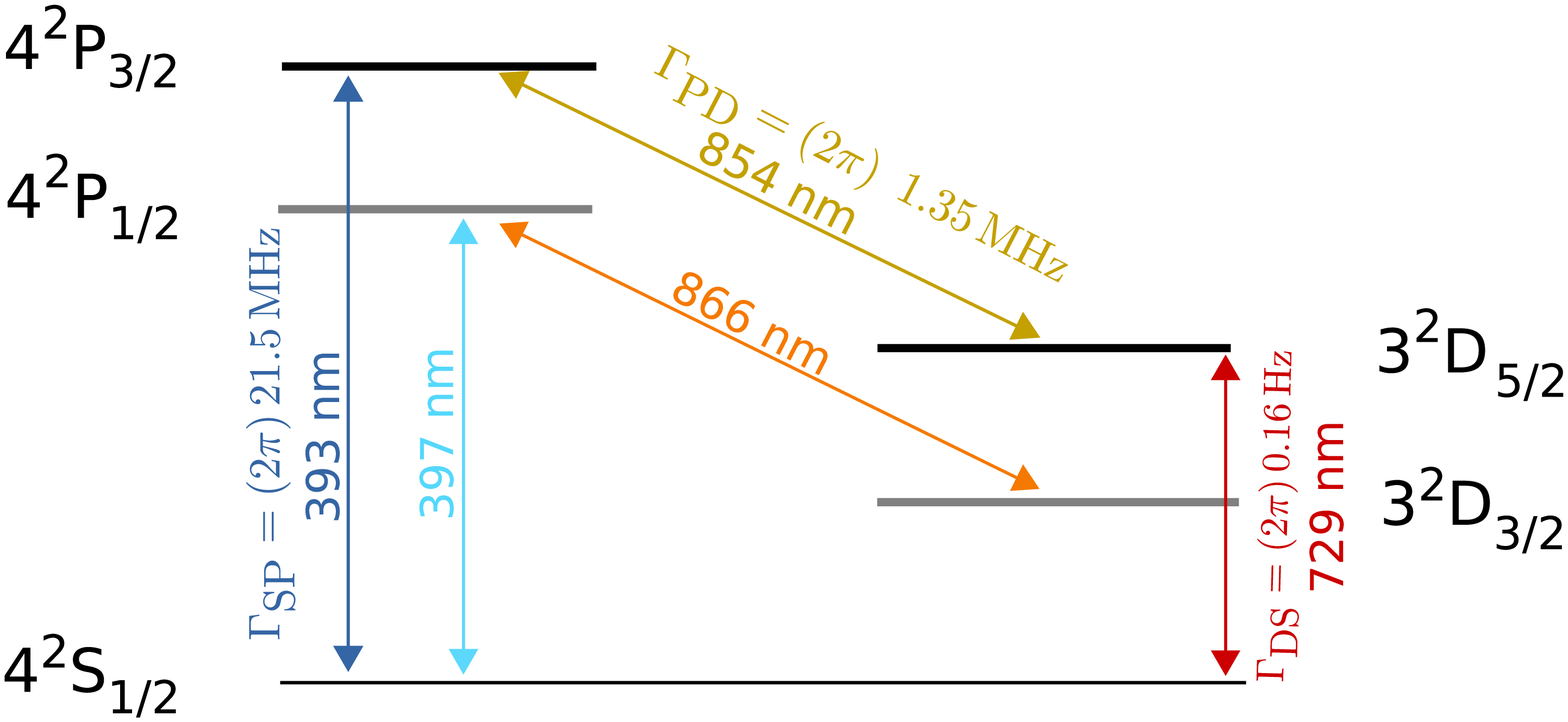}}
  \caption{Simplified level scheme of \Ca40 with relevant transitions and their wavelengths and linewidths. The cavity is on resonance with the transition at 854~nm. The 393~nm transition is driven in order to generate photons in the cavity via a Raman process. The quadrupole transition at 729~nm is used for coherent manipulation of the atomic qubit. Doppler cooling is provided on the transition at 397~nm. The 854~nm and 866~nm transitions are used for repumping.}
  \label{fig_Ca_scheme}     
\end{figure}

The realization of an atom-photon interface requires a coherent mapping between an atomic and a photonic qubit. As a first step, we 
identify quantum states and transitions in the \Ca40 ion suitable on the one hand for initialization, manipulation and detection of the atomic qubit and on the other for mapping between the atomic and the photonic qubit. Because this mapping needs to be reversible, it is advantageous to couple one of the atomic transitions to the mode of an optical resonator \cite{Cirac97}. In this section, we first discuss possible transitions for this coupling, through which single photons are generated in the cavity, and we then treat the definition and manipulation of the atomic qubit.

The relevant \Ca40 level structure is shown in Fig.~\ref{fig_Ca_scheme}. It consists of one ground state manifold, \ssoh; two excited state manifolds, \spoh and \spth; and two metastable state manifolds, \sdth and \sdfh. Within this level structure, a resonator could be coupled either to one of the dipole transitions, \mstmp or \mptmd, or to the quadrupole-allowed transitions, \mstmd. The strongest coupling would be achieved by using the \mstmp transitions, at 397 or 393~nm; however, engineering highly reflective mirror coatings at ultraviolet wavelengths is challenging. On one of the \mstmd transitions, a resonator has been coupled to a single ion with a rate $g = 2 \pi \times 134$~Hz, three orders of magnitude larger than
the atomic decay rate $\gamma_\mathrm{SD} = 2 \pi \times 0.07$~Hz \cite{Mundt02}. Nevertheless, for single photons generated on these time scales, the coherence of the ion-cavity coupling would be obscured by technical imperfections such as magnetic field fluctuations and laser instability. In contrast, two of the \mptmd transitions offer both a strong transition strength ($\gamma_\mathrm{PD} \sim$ 1~MHz) and a wavelength ($\lambda =$ 866, 854~nm) at which ultralow-loss mirror coatings are possible. This transition thus emerges as the most attractive for an atom-photon interface. 

Quantum state transfer as described in Ref. \cite{Cirac97} requires two distinct ground states, coupled via a cavity-assisted Raman process. When the cavity is resonant with the \mptmd transition, one ground state lies in the metastable \md manifold. The second ground state could also be selected from this manifold as in Ref. \cite{Herskind09}, but instead we choose the ground state from the \ms manifold, as in 
Refs. \cite{Guthoehrlein01,Russo09}.
The atomic qubit is thus comprised of one \ms and one \md state. 

In contrast to Refs. \cite{Guthoehrlein01,Russo09}, by using a state from the \sdfh rather than the \sdth manifold, we can take advantage of the tools for initialization, coherent manipulation and state detection developed in the context of quantum information processing \cite{Haeffner08}.
Laser pulses on the \ssohtsdfh transition at 729~nm between Zeeman sub-states can be used to carry out unitary transformations within this qubit space, for example, to generate an arbitrary superposition of the two qubit states or to rotate the measurement basis. In order to detect the qubit state, we make use of the fluorescence detection method, in which lasers tuned to both \ssohtspoh and \spohtsdth transitions excite the atom. Since the \spoh state does not decay to the \sdfh state, the \sdfh state remains uncoupled to any laser during detection, and one can distinguish between the two qubit states by measuring the atomic fluorescence rate.

\subsection{Selection of Zeeman states} \label{raman_transitions_in_ca}
\begin{figure}
\centering
  \resizebox{0.48\textwidth}{!}{
    \includegraphics[width=0.48\textwidth]{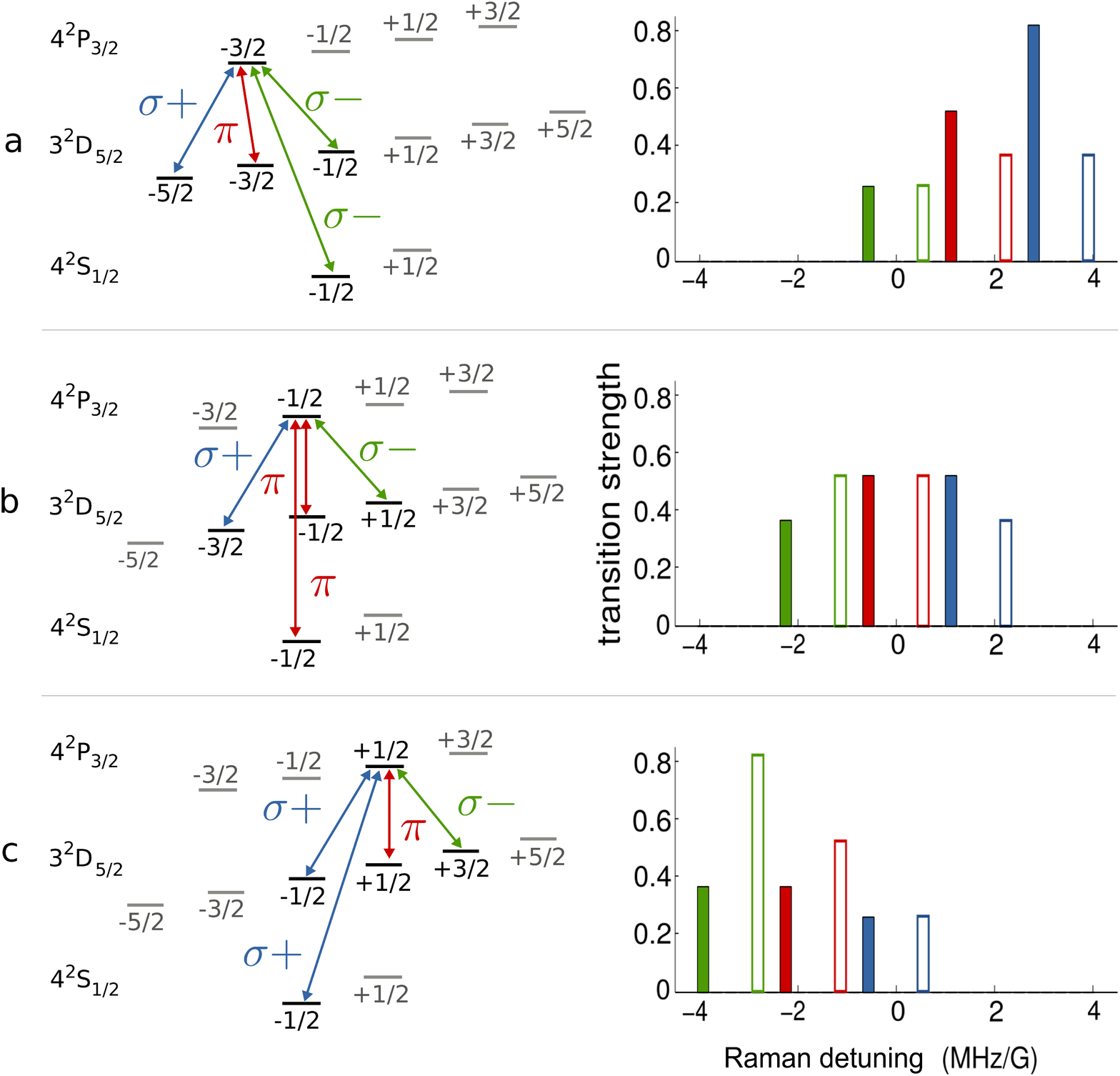}}
  \caption{\ramantrans Raman transition strengths. \textbf{(Left)} Simplified level scheme showing all electronic states available for Raman transitions at nonzero magnetic field. We consider three drive laser polarizations, \textbf{(a)} $\sigma^-$, \textbf{(b)} $\pi$ and \textbf{(c)} $\sigma^+$. For each drive laser polarization, three Raman transitions are possible from each \ssoh state. The polarization of the photons emitted on these transitions are $\sigma^+, \pi$ and $\sigma^-$. \textbf{(Right)} Schematic plot of the strengths of the Raman transitions corresponding to these drive laser polarizations. The filled (unfilled) bars represent the transitions that couple to the \sohmmoh (\sohmpoh) state. The Raman detuning is the detuning of the drive laser or the cavity at nonzero magnetic field. Note: the relative transition strengths in the experiment also depend on the projection of the emitted photons' polarization onto the plane in which cavity photons are polarized.}
  \label{fig_393_854_raman_transitions}     
\end{figure}
Each of the manifolds involved in the \ramantrans Raman transition offers a rich Zeeman structure, which can be exploited for realizing quantum interface schemes. For the purpose of encoding a photonic qubit in its polarization degree of freedom, one can imagine a model in which two and only two cavity-assisted Raman transitions are possible. That is, the two transition paths could either share an initial state but not a final state, or they could connect different initial states to the same final state. In the first case, we identify the two final states as an atomic qubit, whereas in the second case, the initial states constitutes the atomic qubit. If these two Raman transitions produce photons with orthogonal polarizations, then each state of the atomic qubit can be identified with one transition and thus one polarization state of the photon. 

In this section, we focus on the case in which two transition paths begin in the same state and identify a favorable experimental configuration. This case corresponds to the generation of atom-photon entanglement. The inverse case, which corresponds to a mapping of the atomic qubit to the photonic one, is not considered here but can be treated analogously.

In \Ca40, starting in one of the \ssoh states, we apply a drive laser on the \ssohtspth transitions in order to turn on the Raman coupling. For every possible polarization ($\sigma^+, \pi$ or $\sigma^-$) of this laser, there exist three Raman transitions, each coupling to a different state of the \sdfh manifold as shown in Fig.~\ref{fig_393_854_raman_transitions}. 
At zero magnetic field, all Zeeman sub-states of one manifold are degenerate, and therefore all Raman transitions overlap in frequency, resulting in an unfavorable situation: three transitions are possible instead of two. The number of allowed Raman transitions thus has to be reduced by one. 

One option to disable one of the three Raman transitions at zero magnetic field is realized in neutral atom experiments \cite{Volz06,Wilk07b}. If the cavity axis is chosen as the quantization axis, we see that the polarization of $\pi$ photons lies along the cavity axis and that therefore only circularly polarized photons are emitted into the cavity. For the case of a drive laser with propagation orthogonal to the cavity axis 
and linear polarization along the cavity axis, the initial state is coupled to only one of the \spth states. As a consequence, only two Raman transitions are driven, which generate $\sigma^+$- and $\sigma^-$-polarized photons in the cavity, as in our model system.

Alternatively, one can apply a magnetic field $B$, which lifts the degeneracy between Zeeman sub-states. Raman transitions are then split by 
\begin{align}
 \Delta E_{{S_{1/2}, m_J}\leftrightarrow{D_{5/2}, m_{J'}}} = - \mu_\mathrm{B} B (g_{D_{5/2}} m_J - g_{S_{1/2}} m_{J'}),
 \label{zeeman_splittings}
\end{align}
where $\mu_\mathrm{B}$ is the Bohr magneton and $g_{L_{J}}$ is the Land\'{e} factor of level $L_{J}$ with orbital angular momentum $L$ and total angular momentum $J$. This splitting enables individual addressing of all Raman transitions via the detuning of the drive laser $\delta_\mathrm{drv}$ or the cavity $\delta_\mathrm{cav}$ and the polarization of the drive laser. Since for an atom-photon interface, it is advantageous to have the polarization and frequency degrees of freedom of the photon uncorrelated, the individual transitions should be addressed by the frequency and polarization of the drive laser only, while the cavity detuning remains fixed. Thus, photons generated by all Raman transitions have the same frequency. 
In this case, our model can be realized by driving two Raman transitions at two distinct drive-laser frequencies simultaneously. For a proper definition of the photonic qubit, two transitions have to be chosen so as to generate photons in orthogonal polarization modes of the cavity. 

As the individual addressing of the \ssohtsdfh transitions required for atomic qubit manipulation and detection relies on a nonzero magnetic field, we opt for the second scenario, as it avoids a change of the magnetic field during the experiment. 
 
Out of the nine \ramantrans Raman transitions available from each \ms state (Fig.~\ref{fig_393_854_raman_transitions}), we now select two according to the following considerations. 
(In the discussion that follows, we assume that the Raman resonance condition is satisfied, that is, that $\delta_\mathrm{drv}$ exactly balances the sum of $\delta_\mathrm{cav}$ and all Zeeman and Stark shifts of the initial and final states.)
First, the coherence of photon generation in our system is determined by the ratio of the effective Raman coupling strength $\Omega_\mathrm{eff}$ to the effective spontaneous emission rate of the atom $\gamma_\mathrm{eff}$ \cite{Barros09}. 
The effective Raman coupling comprised of the \ssohtspth transition $i$ and the \spthtsdfh transition $j$ has an amplitude of
\begin{eqnarray}
  \Omega_{\mathrm{eff},ij} \xspace \approx \xspace \dfrac{\alpha_i \; \Omega_\mathrm{drv} \cdot \beta_j \; 2 g_0}{2 |\delta_\mathrm{drv}|} \mathrm{,}
\end{eqnarray} 
where $\Omega_\mathrm{drv}$ is the drive-laser Rabi frequency, $\delta_\mathrm{drv}$ is the drive-laser detuning, and $g_0$ is the strength of the ion-cavity coupling to the \spthtsdfh transition. The coefficient $\alpha_i$ is the product of the projection of the drive laser polarization onto the dipole moment of transition $i$ with the Clebsch-Gordon coefficient of this transition. Analogously, $\beta_j$ is the product of the projection of the polarization plane of the cavity mode to the atomic dipole moment of transition $j$ with the Clebsch-Gordon coefficient of this transition. The effective spontaneous emission rate of the atom is given by
\begin{eqnarray}
 \gamma_\mathrm{eff} \approx \gamma \left( \dfrac{\Omega_\mathrm{drv}}{2 |\delta_\mathrm{drv}|} \right)^2 \mathrm{,}
\end{eqnarray}
which becomes independent of transition $i,j$ when the detuning $\delta_\mathrm{drv}$ is much larger than the Zeeman splittings $\Delta E$ given by Eq. \ref{zeeman_splittings}. For a fixed ratio $\Omega_\mathrm{drv}/|\delta_\mathrm{drv}|$ and $g_0$ determined by the cavity geometry, in order to maximize the ratio $\Omega_\mathrm{eff}/\gamma_\mathrm{eff}$, one should maximize the product $\alpha_i \cdot \beta_j$. 

A second criterion for selection is that the two transitions have similar strengths, since then mapping of both components of an arbitrary atomic superposition to the photonic one occurs at the same rate. 

In order to meet both criteria, i.e., both high and similar coupling strengths, two scenarios emerge as attractive. 
In the first scenario, the magnetic field is oriented along the cavity axis. If we choose the magnetic field axis as the quantization axis, the 
possible polarization states of photons in the cavity are then $\sigma^+$ and $\sigma^-$. If we choose the initial state to be \fm{{}\ket{\mathrm{S}_{1/2}, m_j = - \frac{1}{2}}}\xspace (in the following denoted as \sohmmoh), then the optimal transition pair is \sohmmoh$\leftrightarrow$\xspace\pthmmoh$\leftrightarrow$\xspace\dfhmmth and \sohmmoh$\leftrightarrow$\xspace\pthmmoh$\leftrightarrow$\xspace\dfhmpoh, with transition strengths $\alpha\cdot\beta=$~(0.52, 0.37). The corresponding transitions from \sohmpoh have identical transition strengths. 

In the second scenario, the direction of the magnetic field is orthogonal to the cavity axis. We again identify the magnetic field axis as the quantization axis; photons emitted by the atom with circular polarization are now projected to horizontally-polarized ($H$) cavity photons, while linearly-polarized $\pi$ photons are projected to vertically-polarized ($V$) cavity photons, where this assignment defines $H$ and $V$. Again, we assume initial state \sohmmoh.  
The optimal transition pair is given by \sohmmoh$\leftrightarrow$\xspace\pthmmth$\leftrightarrow$\xspace\dfhmmfh and \sohmmoh$\leftrightarrow$\xspace\pthmmth$\leftrightarrow$\xspace\dfhmmth, with strengths $\alpha\cdot\beta=$~(0.58, 0.52), where the drive beam is circularly polarized. (The corresponding transitions from \sohmpoh again have identical strengths.)  As these transition strengths are both larger and more similar than in the first scenario, this pair is the most suitable for the generation of atom-photon entanglement.

\section{Characterization of the setup} \label{setup}
  We summarize the relevant parameters of the experimental setup, previously described in 
detail in Ref. \cite{Russo09}, and highlight recent additions. 

\subsection{Experimental apparatus}
\begin{figure}
  \resizebox{0.48\textwidth}{!}{%
    \includegraphics[width=0.48\textwidth]{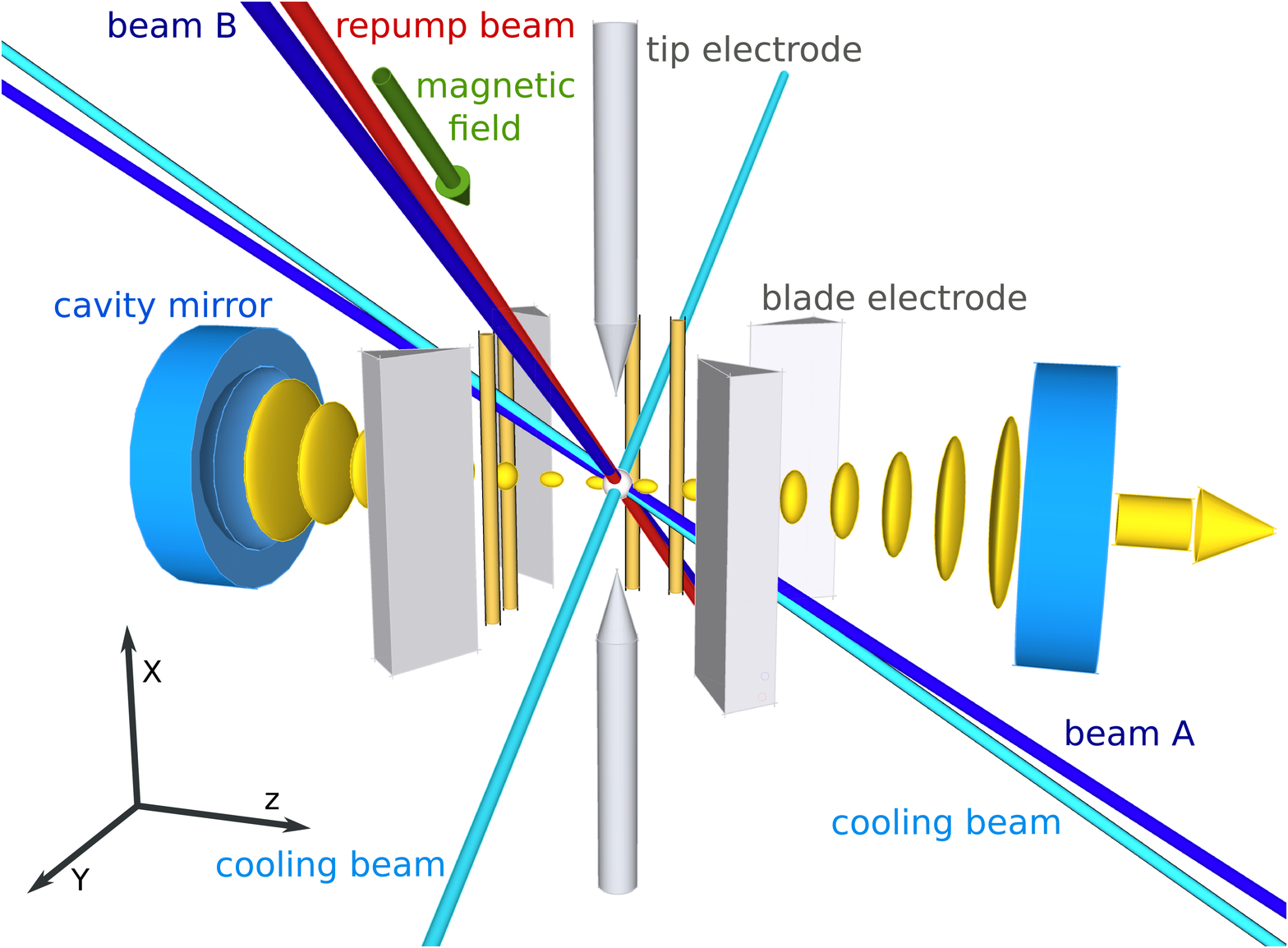}}
  \caption{Trap and cavity geometry: the ion trap consists of tip and blade electrodes and four additional electrodes for the compensation of micromotion (yellow). The cavity consists of two mirrors mounted around the trap and confines a standing wave; an arrow at one mirror indicates the output channel. The orientation of the magnetic field is perpendicular to the cavity axis. The Raman transition is driven by beam A (linearly polarized) or beam B (circularly polarized), both at 393~nm. Two beams at 397~nm cool the ion. Repump beams at 854~nm and 866~nm share the same path. Not shown is the laser beam at 729~nm.}
  \label{3Dgeometry_of_setup}     
\end{figure}

A single \Ca40 ion is stored in a linear Paul trap with trapping frequencies of $\nu_\mathrm{a}= 2 \pi \times 1.1$~MHz along the trap axis of symmetry and $\nu_\mathrm{r1,r2} = 2\pi \times (3.00, 3.05)$~MHz in the radial plane. In order to perform Doppler cooling and state detection of the ion, laser beams near resonance drive the \ssohtspoh transition at 397~nm and the \spohtsdth transition at 866~nm; an additional 397~nm beam has been introduced into the setup of \cite{Russo09} for improved Doppler cooling (Fig.~\ref{3Dgeometry_of_setup}). In the remainder of the text, these beams are referred to as the cooling laser and repump laser, respectively. We detect fluorescence light on the \ssohtspoh transition with both a photomultiplier tube (PMT) and a CCD camera. 

In order to initialize, coherently manipulate and detect the state of the atomic qubit, we now drive the quadrupole \ssohtsdfh transition at 729~nm with a tapered-amplifier diode laser seeded by a Ti:Sapphire laser stabilized to a high-finesse cavity \cite{Chwalla09}, with a linewidth broadened to $\approx$ 0.4 kHz by fiber noise.  
This laser is also used to characterize trap frequencies, magnetic fields at the ion position, and heating rates and micromotion of the ion. Moreover, the laser enables sideband cooling \cite{Leibfried03} in all three dimensions, as demonstrated in Section \ref{results}.

We have introduced a frequency-doubled Ti:Sapphire laser with a linewidth of $\approx$ 30 kHz in order to drive the \ssohtspth transition at 393~nm, and thus the cavity-assisted Raman transitions to the \sdfh manifold identified in Section \ref{transitions}. This drive laser can be sent to the ion along two paths, which are indicated in Fig.~\ref{3Dgeometry_of_setup}: beam A propagates at an angle of $\sim$45\textdegree~to both the cavity and trap axis with linear polarization orthogonal to the magnetic field; beam B propagates in the direction of the magnetic field with circular polarization.   Beam B can thus be used to enact the second ion-photon entanglement scheme proposed at the end of Section \ref{raman_transitions_in_ca}.  The ion is repumped on the \spthtsdfh transition by a diode laser at 854~nm linearly polarized orthogonal to the magnetic field, with secondary repumping provided by the 866 nm laser.

Note that the coordinate system in Figs. \ref{3Dgeometry_of_setup} and \ref{fig_standing_wave} has been chosen as a natural basis in which to describe the relative positions of ion and cavity. It does not refer to the quantization axis of the ion.

Two mirrors with radius of curvature $R=$ 10 mm constitute a near-concentric optical resonator around the trap. Due to asymmetric mirror transmissions $T_1 = 1.3(3)$ ppm and $T_2 = 13(1)$ ppm, the cavity field decays preferentially through one mirror of the resonator. At the cavity wavelength $\lambda =$ 854~nm, resonant with the \spthtsdfh transition, we measure a decay rate of $\kappa = 2 \pi \times 50$~kHz. After traversing a polarization analysis setup consisting of a half- and a quarter-waveplate and a polarizing beamsplitter (PBS) \cite{James01}, photons are detected by two avalanche photodiodes (APDs) with dark count rates of (33.1, 33.6)~Hz and detection efficiencies of ($49\pm4$, $46\pm4$)\%. The overall probability with which a photon in the cavity will be detected at the APDs is ($8.1\pm1.5$, $7.6\pm1.5$)\%; this probability is given by the product of the APD efficiencies with the optical path transmissions  ($87\pm3$, $86\pm3$)\% and the cavity output coupling efficiency $19\pm3$\%.
Each cavity mirror is mounted on a shear-mode piezo stack. The length of the cavity is actively stabilized with respect to an ultra-stable passive cavity by means of a transfer lock at 783 nm.

The ion-cavity coupling strength is given by 
$g_0 = \sqrt{\frac{3 c \gamma \lambda^2}{\pi^2 L\omega_0^2}}$, where  
$2\gamma=2\gamma_\mathrm{\spth\fm{\rightarrow}\xspace\sdfh}$ is the spontaneous emission rate, $L$ is the length of the cavity and $\omega_0$ is the waist of the TEM$_{00}$ cavity mode. This coupling strength is thus determined by the geometry of the resonator. 
We measure the free spectral range of the cavity in order to determine $L= (19.96\pm0.02)$ mm.  To estimate the error on the manufacturer-specified value of $R$, we measure the frequency splitting of the TEM$_{00}$ and TEM$_{01}$ modes, which depends on both $L$ and $R$. We infer $R=$ (10.02$\pm$0.01) mm, from which we calculate $\omega_0 = \sqrt{\frac{\lambda}{2\pi} (L(2R-L))^{1/2}}= (13.2\pm0.8)$ \mum and thus a maximum coupling rate of $g_0 = 2 \pi \times (1.43 \pm 0.01)$~MHz for the \sdfhtspth transition.

\subsection{Ion-cavity localization}
\begin{figure}
\centering
  \resizebox{0.48\textwidth}{!}{%
    \includegraphics[width=0.48\textwidth]{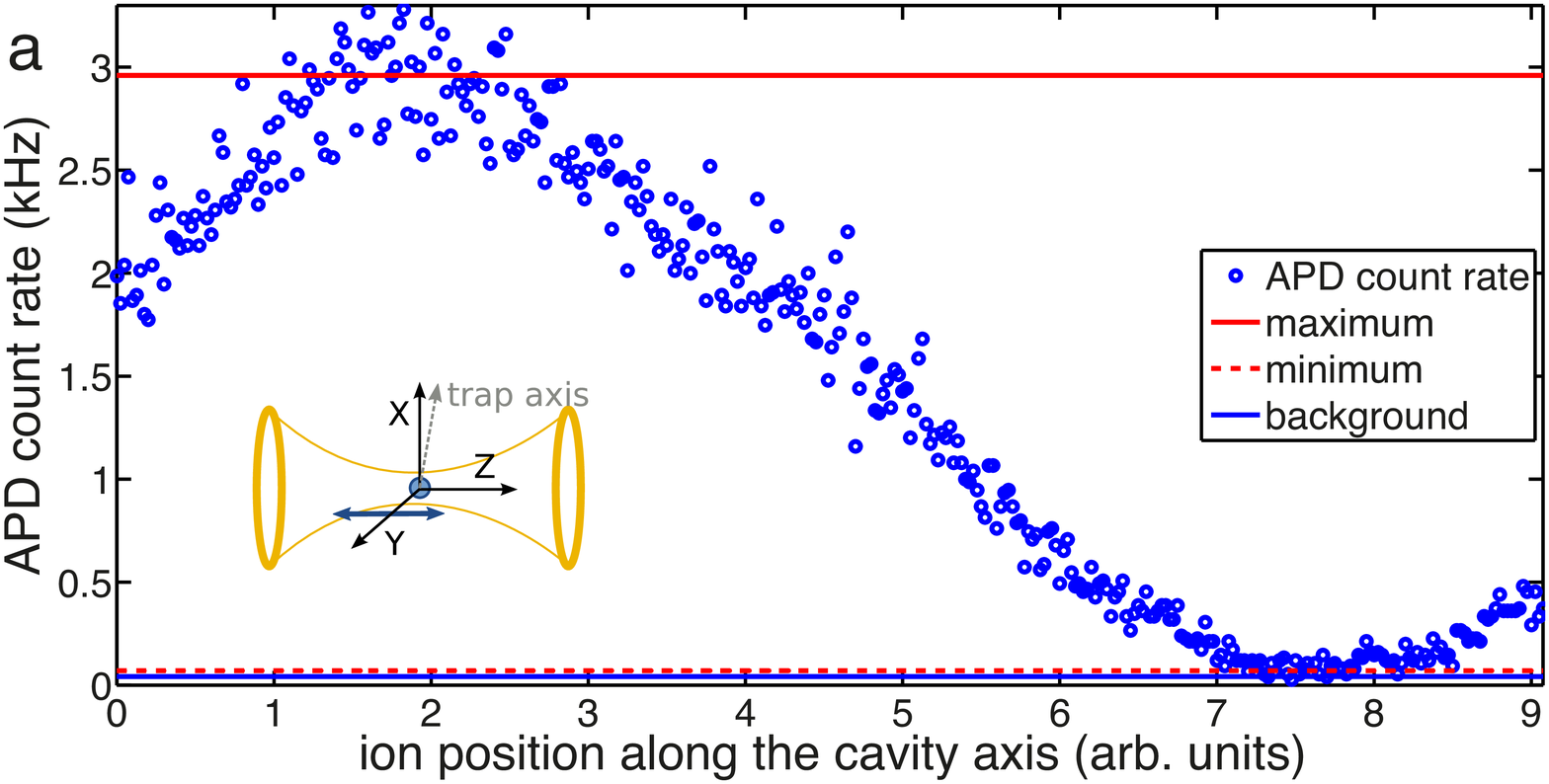}}
  \resizebox{0.48\textwidth}{!}{%
  \includegraphics[width=0.48\textwidth]{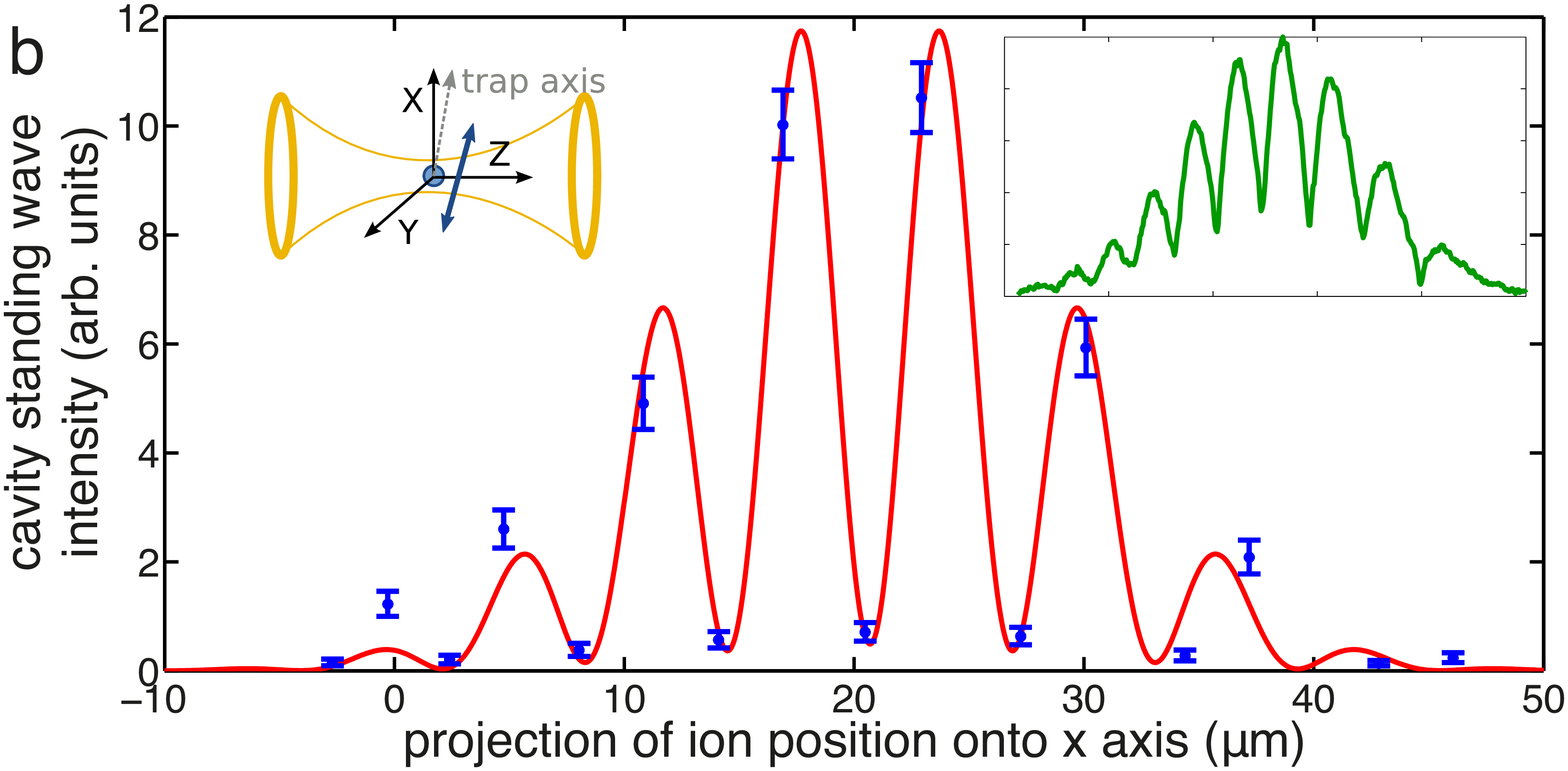}}
  \caption{a) Translation of the cavity standing wave with respect to the ion.  The ion-cavity system is driven on Raman resonance with the cavity, 400 MHz red-detuned from the transition at 393 nm; the number of 854 nm photons generated in the cavity depends on the ion-cavity coupling, which varies from maximum to minimum across the standing wave.  The measurement sequence described in the text is recorded 250 times for each data point. b) Translation of the ion along the trap axis of symmetry, which is nearly orthogonal to the cavity.  The ion is driven with the cooling beam near resonance at 397 nm and is repumped by a classical standing wave within the cavity at 866 nm.  The Gaussian envelope of the intensity at 866 nm, extracted from the resonance fluorescence, is determined by the convolution of the cavity waist $\omega_0$ and relative ion-cavity motion.  The inset shows a full scan for a fixed position of the cooling beam, while for the data in the central plot, the cooling beam is re-centered on the ion at each data point. (The inset displays the count rate on the CCD camera vs. the ion position in arbitrary units.)}  
  \label{fig_standing_wave}     
\end{figure}
Experimentally, one observes a value of the coupling strength $g_\mathrm{obs} \leqslant g_0$ depending on the position of the ion with respect to the standing wave of the resonator \cite{Mundt02}. We are able to adjust this relative position by means of in-vacuum piezo stages and a feedthrough \cite{Russo09}. However, the position may vary on fast time scales 
due to both the temperature of the ion in the trap and center-of-mass vibrations of the cavity with respect to the trap. These relative vibrations can occur because cavity and trap are mounted separately and rigidly to the vacuum chamber \cite{Russo08}. 
In order to quantify the extent of relative motion, we perform two measurements. 

In the first measurement, we observe the localization of the ion in the standing wave of the cavity by translating the cavity with respect to the ion. 
We tune the drive laser and the frequency-stabilized cavity to satisfy the Raman resonance condition, 400 MHz red-detuned from the \spth excited state. After 2~ms of Doppler cooling the single ion, we apply the drive and repump laser simultaneously for 300 \mus while detecting cavity photons at the APDs. By scanning the voltage on one cavity mirror's piezo stack  while sending a Pound-Drever-Hall feedback signal to the other stack, we shift the cavity along its axis while keeping its length fixed. The range of this piezo stack is approximately one cavity standing wave.  
As the position of the ion moves from an antinode to a node of the cavity standing wave, the Raman coupling between the \ms and \md states and thus the number of photons in the cavity changes from maximum to minimum, as seen in Fig.~\ref{fig_standing_wave}a. The visibility of this modulation, observed on the APDs, is ($98 \pm 2$)~\%, corresponding to a localization of ($13 \pm 7$)~nm \cite{Eschner03}. 

In comparison to our previously measured visibility of 60\% \cite{Russo09}, we have thus been able to decrease the residual motion of the ion along the cavity axis to the point where our minimum signal is just above background. This improvement was accomplished by optimizing Doppler cooling. From independent measurements at 729 nm, we have determined that cooling via one laser beam at 397~nm results in unequal cooling of the two radial modes and temperatures well above the Doppler limit. We now apply two beams simultaneously from different directions as shown in Fig.~\ref{3Dgeometry_of_setup}, providing uniform cooling near the Doppler limit in all three dimensions. These two beams have orthogonal polarizations in order to avoid interference effects at the position of the ion.
As the localization is also on the order of the Doppler-cooling limit,
we conclude that any relative motion of the cavity with respect to the trap in the direction of the standing wave is negligible.

In the second measurement, we translate the ion by changing the relative voltage of the tip electrodes, which shifts the trap minimum along the trap axis of symmetry. In this way, we can probe the radial structure of the TEM$_{00}$ cavity mode, which we expect to have waist $\omega_0 = (13.2 \pm 0.8)$ \mum. If relative motion orthogonal to the cavity on the \mum scale  
is present, we expect to measure a broadened waist, representing the convolution of the motion and the mode structure.

We drive the cavity with the repump laser at 866 nm in order to generate a classical standing wave field.  The ion is then driven near resonance at 397 nm with the cooling laser.  Because the repump is below saturation, ion fluorescence is position-dependent and can be used to extract the 866 nm intensity as a function of position.  Both fluorescence and position are measured on the CCD camera.  The resulting position-dependent intensity reveals a periodically
modulated Gaussian mode (right inset in Fig.~\ref{fig_standing_wave}b) instead of the simple Gaussian that one expects. 

We can explain this modulation by assuming that the trap axis of symmetry is not completely orthogonal to the cavity, due to imperfect alignment in assembly of the experiment. Therefore, as the ion is translated along the trap axis, it also intersects the standing wave of the cavity mode. 
From the number of fringes visible and the wavelength of the standing wave field, we extract the deviation of this angle from perpendicular to be 4\textdegree. (We observe a similar structure when we tune the relative ion-cavity position along the axis perpendicular to both cavity and trap, but we are unable to calibrate the length scale in this direction.) We note that these deviations do not affect the cavity-QED experiments described in this text, which are carried out for a fixed ion position. 

For the data shown in the right inset of Fig.~\ref{fig_standing_wave}b, the ion is continuously displaced by 60~$\mu$m along the trap axis.  Although the periodic intensity modulation is visible, the intensity of the cooling beam varies significantly along the path of the ion, resulting in a nonuniform ion temperature.  In a more accurate measurement, shown in the main part of Fig.~\ref{fig_standing_wave}b, we align the cooling beam to the ion position for every data point, recorded at maxima and minima of the modulated Gaussian field.

In order to extract any motion $\sigma_{x,y,z}$ which broadens our measured waist, we model the structure of the data in Fig.~\ref{fig_standing_wave}b by assuming a Gaussian localization of the ion wavepacket \begin{eqnarray}
\vert \psi(x,y,z) \vert^2 = \frac{1}{(2\pi)^{3/2}\sigma_x \sigma_y\sigma_z}e^{\frac{-x^2}{2\sigma_x^2}}e^{\frac{-y^2}{2\sigma_y2}} e^{\frac{-z^2}{2\sigma_z^2}}.
\end{eqnarray}
We define a coordinate system in which the $z$-axis is oriented along the cavity axis and the trap axis lies in the $xz$-plane (i.e., the trap axis is situated at an angle of 4\textdegree to the $x$-axis in this plane).  
The intensity of the cavity field is given by 
\begin{eqnarray}
I(x,y,z) = I_{0} e^{\frac{-2x^2}{\omega_0^2}} e^{\frac{-2y^2}{\omega_0^2}} \sin^2(\frac{2 \pi}{\lambda} z),
\end{eqnarray}
where $\lambda =$ 866~nm.  We have approximated the waist of the TEM$_{00}$ mode as $\omega_0$, since the range of travel in the $z$ direction is much smaller than the Raleigh range of $640$~\mum. The expected intensity profile $I_\mathrm{eff}$ seen by the ion is then given by the convolution of $\vert \psi(x,y,z) \vert^2$ with $I(x,y,z)$.
We solve for $I_\mathrm{eff}(x,y,z)$ analytically and set $y=0, z=x\tan{4^\circ}$ to parameterize $I$ as a function of $x$, as in the measurement of Fig.~\ref{fig_standing_wave}b: 
\begin{eqnarray}
I(x)\propto e^{-2x^2/(4\sigma_x^2+\omega_0^2)} (1-\cos{\frac{4\pi x \tan{4^\circ}}{\lambda}}e^{-8\pi^2 \sigma_z^2/\lambda^2}).
\end{eqnarray}
We fit this function to the data of Fig.~\ref{fig_standing_wave}b and obtain values $\sigma_x = 4.7\pm2.2$~\mum and $\sigma_z = 48\pm46$~nm. 
The value for $\sigma_z$ is consistent with the localization of the previous measurement.  It is surprising that micron-scale motion is only orthogonal to the cavity axis; we hypothesize that this motion is due to vibration along the axis of the cavity mount, coupled into the chamber via the translational feedthrough.

The effect of $\sigma_x$ on the ion-cavity coupling is given by
\begin{align}
 g_{\mathrm{obs}} &=  \frac{1}{\sqrt{2\pi}\sigma_x} \int_{-\infty}^{\infty} dx~e^{\frac{-x^2}{2\sigma_x^2}} g_0 e^{\frac{-x^2}{\omega_0^2}}
=  (0.89\pm0.06)g_0. 
 \end{align}

Thus, the observed coupling along the $x$ axis is only slightly reduced from its maximal value, and we have shown in the previous measurement that a reduction along the $z$ axis is negligible.  Although we are not able to quantify the extent of motion along the $y$ axis, the agreement of our data with simulations in which $g = g_{\mathrm{obs}}$ (Section \ref{results}) suggests that it does not contribute significantly.

For experiments involving two ions, the observed modulation in the coupling along the trap axis allows for precise control of the ions' individual coupling to the mode of the cavity. As the typical spacing between two ions is on the order of $\sim$ 5 \mum, we have the ability to place the ions at neighboring antinodes by adjusting the voltage of the tip electrodes; in this case, both ions couple with similar, near-maximal strength. Conversely, we can position one ion at a node and the other at an antinode, thus coupling only one ion to the cavity.

\section{Raman spectroscopy, orthogonal photons and coherent state manipulation} \label{results}
  Having established that the coupling of an ion to our cavity mode is not significantly limited by relative motion, we now demonstrate key building blocks of an ion-cavity interface.  We identify our target Raman transitions via spectroscopy, generate orthogonal single photons on these two transitions, and analyze coherent manipulation of the associated atomic states.  

\subsection{Raman spectroscopy}
\begin{figure}
\centering
  \resizebox{0.5\textwidth}{!}{%
    \includegraphics{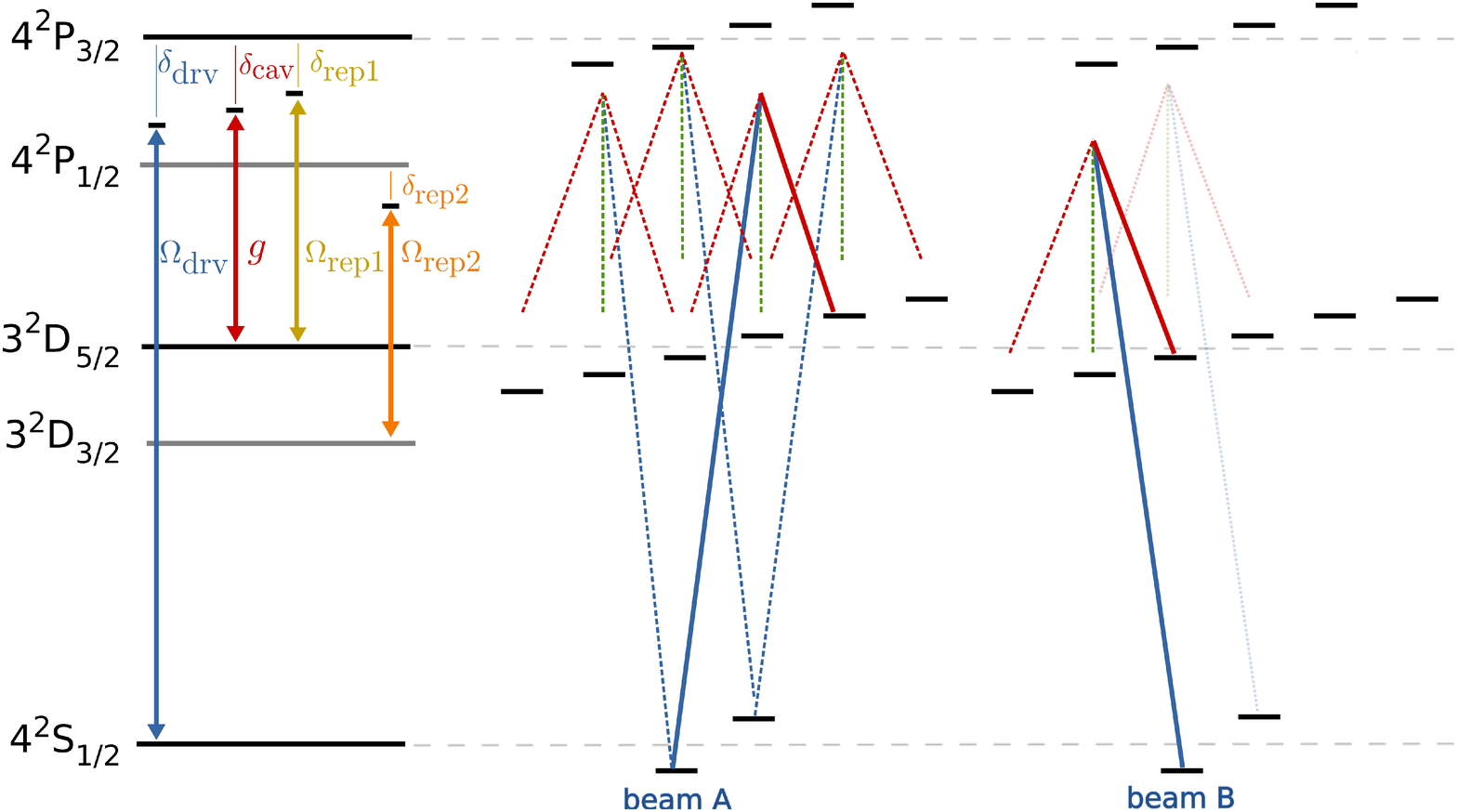} }
  \caption{Driving schemes for the \ramantrans Raman transitions. \textbf{(Left)} Simplified level scheme of \Ca40 with relevant ion-field couplings: $\Omega_\mathrm{drv}, \Omega_\mathrm{rep1}, \Omega_\mathrm{rep2}$ are the Rabi frequencies of the drive and the two repump lasers, respectively, and $\delta_\mathrm{drv}, \delta_\mathrm{rep1}, \delta_\mathrm{rep2}$ are the detunings of these lasers from resonance. The Rabi frequency of the ion-cavity coupling is $2g$, where the cavity is detuned by $\delta_\mathrm{cav}$ from the \sdfhtspth transition. \textbf{(Center)} 
 Due to its linear polarization orthogonal to the magnetic field, beam A drives $\sigma^{+}$ and $\sigma^{-}$ transitions (blue arrows). Photons emitted into the cavity have horizontal (red) or vertical polarization (green). For the value of $\delta_\mathrm{drv}$ indicated here, only the transition illustrated with a solid line is resonant. \textbf{(Right)} 
Beam B has $\sigma^{-}$ polarization and thus provides optical pumping, suppressing three of the six possible Raman transitions (transparent lines).}
  \label{fig_continuous_driving}     
\end{figure}

In order to locate specific cavity-mediated Raman transitions, we first probe the spectrum of Raman transitions between the \ssoh and \sdfh manifolds by scanning the detuning of the drive laser $\delta_\mathrm{drv}$; the cavity detuning $\delta_\mathrm{cav}\approx 2 \pi \times 400$~MHz remains fixed.

The experimental sequence is as follows: after Doppler cooling the ion for 2~ms, we apply the drive laser simultaneously with the near-resonant 854~nm and 866~nm repump lasers for 300 \mus. During this interval, we record photons emitted from the cavity at the APDs, where the waveplates in front of the PBS at the cavity output have been set to measure photons in the ($H,V$) basis. We repeat this sequence 250 times for each value of $\delta_\mathrm{drv}$. The observed spectrum depends on the polarization of the drive laser beam.   The frequencies of spectrum peaks are Stark-shifted by the drive laser field.  Rabi frequencies $\Omega_\mathrm{drv}$ are thus calibrated by measuring the spectrum frequency shift due to a known fractional change in the drive intensity.

\begin{figure*}
  \resizebox{0.99\textwidth}{!}{%
    \includegraphics[width=0.99\textwidth]{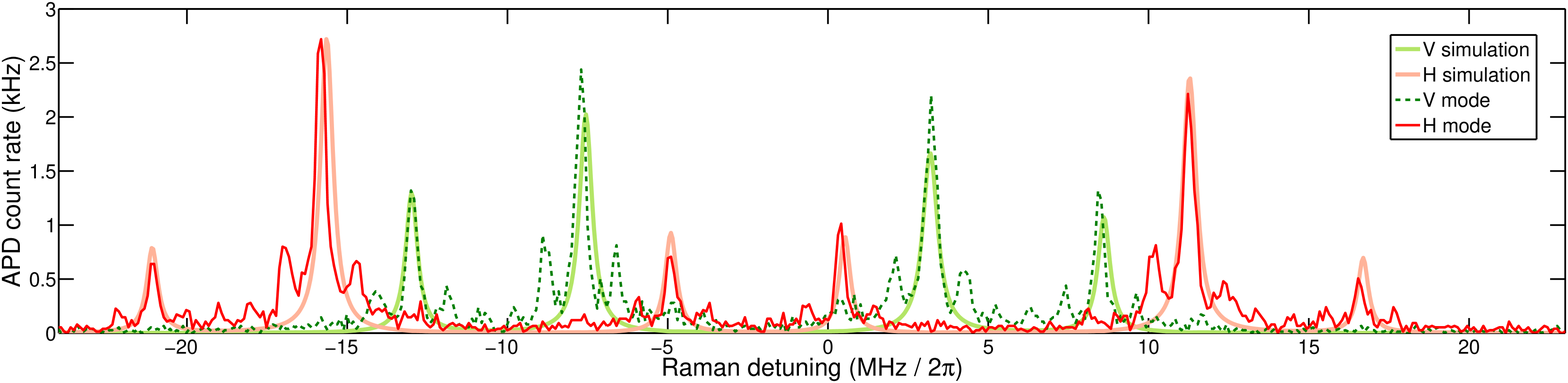} }
  \caption{Raman spectroscopy of the \ramantrans transition.  One arm of the Raman transition is provided by the cavity; the other arm is provided by drive beam A, with $\sigma^+/\sigma^-$ polarization. APD count rates are plotted as a function of the drive laser detuning; photons from the $H$ (red) and $V$ (green) cavity modes are recorded on separate APDs.  A steady-state solution of the master equation, taking into account both cavity modes and 18 levels of \Ca40, agrees well with the data.  The master-equation simulation does not include motion of the ion and thus does not reproduce the sidebands that are visible at secular frequencies $\nu_{\mathrm{a}}$ and $\nu_{\mathrm{r1,r2}}$.}
  \label{fig_raman_spectrum}     
\end{figure*}

\begin{figure}
  \resizebox{0.48\textwidth}{!}{%
    \includegraphics[width=0.48\textwidth]{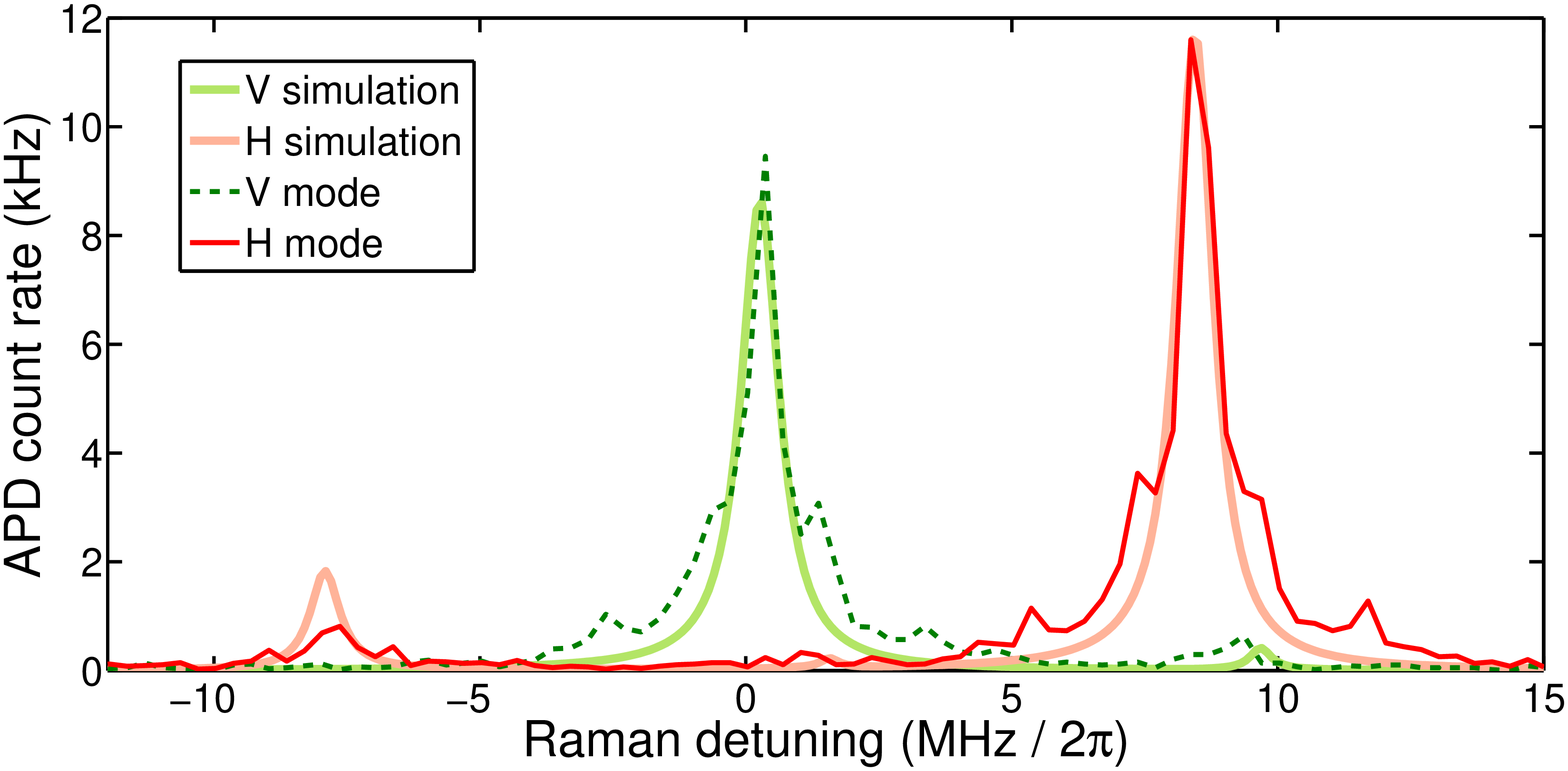}}
  \caption{Raman spectroscopy and master-equation simulation of the \ssohtsdfh transition as in Fig. \ref{fig_raman_spectrum}, but with the Raman transition driven by beam B, which has $\sigma^-$ polarization.}
  \label{fig_raman_spectrum_sigma}     
\end{figure}

Beam A drives the \ssohtspth transition with $\sigma^{+}$ and $\sigma^{-}$ polarization, resulting in 12 possible Raman transitions, indicated in the center of Fig.~\ref{fig_continuous_driving}. However, transitions which share the same initial and final states but are driven via a different virtual state of the \spth manifold are resonant at the same drive laser frequency. We therefore expect ten peaks in the spectrum, all of which are identified in Fig.~\ref{fig_raman_spectrum}. Moreover, for each peak, we are able to resolve sidebands corresponding to the ion's secular motion at frequencies $\nu_{\mathrm{a}}$ and $\nu_{\mathrm{r1,r2}}$. The Rabi frequency of the drive laser is given by $\Omega_\mathrm{drv} = 2 \pi \times 88$~MHz.  This frequency corresponds to values $\Omega_{\mathrm{eff,}i,j} = 2 \pi \times \alpha_{i} \beta_{j } \times 0.31$~MHz and $\Gamma_{\mathrm{eff}} =  2 \pi \times 0.25$~MHz.

Beam B propagates along the magnetic field axis with $\sigma^{-}$ polarization, driving six Raman transitions, indicated in the right part of Fig.~\ref{fig_continuous_driving}. Due to its polarization, beam B depopulates the \sohmpoh state via optical pumping. Although the state can be repopulated by the repumping beams, its steady-state population in simulations is on the order of $5\%$. We thus expect three central peaks in the Raman spectrum, which can be identified in Fig.~\ref{fig_raman_spectrum_sigma} ($\Omega_\mathrm{drv} = 2 \pi \times 99$~MHz, 
$\Omega_{\mathrm{eff,}i,j} = 2 \pi \times \alpha_{i} \beta_{j } \times 0.35$~MHz, $\Gamma_{\mathrm{eff}} =  2 \pi \times 0.32$~MHz). The difference in maximum count rate between the spectra of beams A and B is due primarily to optical pumping: the Raman transition and the repump laser drive a nearly closed cycle, efficiently generating cavity photons. The two strongest transitions driven by beam B are the ones selected in Section \ref{raman_transitions_in_ca} as optimal for atom-photon entanglement.

For both drive beams, the relative heights of the spectrum peaks correspond roughly to the calculated transition strengths (Section \ref{transitions}). The full dynamics of the 18-level system with two orthogonal cavity modes, driven by three lasers, are described by master-equation simulations, also plotted in Fig.~\ref{fig_raman_spectrum} and Fig.~\ref{fig_raman_spectrum_sigma}. For example, one would expect the spectrum of Fig.~\ref{fig_raman_spectrum} to be symmetric, as corresponding transitions from the two \ssoh states have equal strengths, but the asymmetry in height from left to right is due to a detuning of the repump laser at 854 nm.  The peak width is primarily determined by the Rabi frequency $\Omega_\mathrm{drv}$; the background signal is entirely due to dark counts of the APD. The frequency splitting between peaks is determined by the magnetic field of 4.77~G, which we have selected in order to avoid overlap of sidebands from neighboring transitions.  The simulation amplitudes correspond to a cavity output path efficiency of $8.0\%$, consistent with the measured path efficiencies of $(8.1 \pm 1.5, 7.6 \pm 1.5)\%$.

\subsection{Motional sidebands} \label{results_motional_sidebands}
\begin{figure}
\centering
\resizebox{0.48\textwidth}{!}{%
  \includegraphics[width=0.48\textwidth]{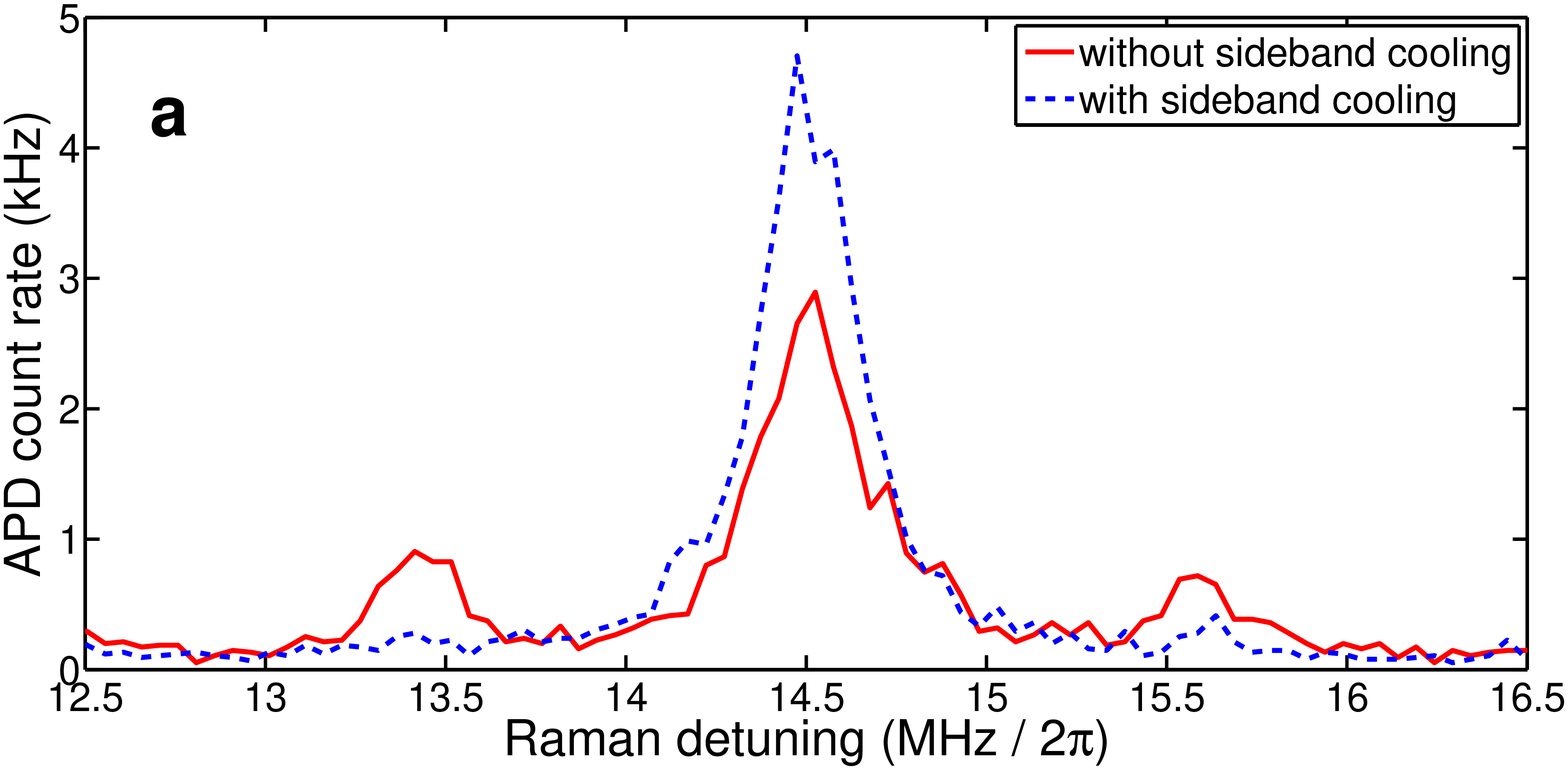}} 
\resizebox{0.48\textwidth}{!}{%
  \includegraphics[width=0.48\textwidth]{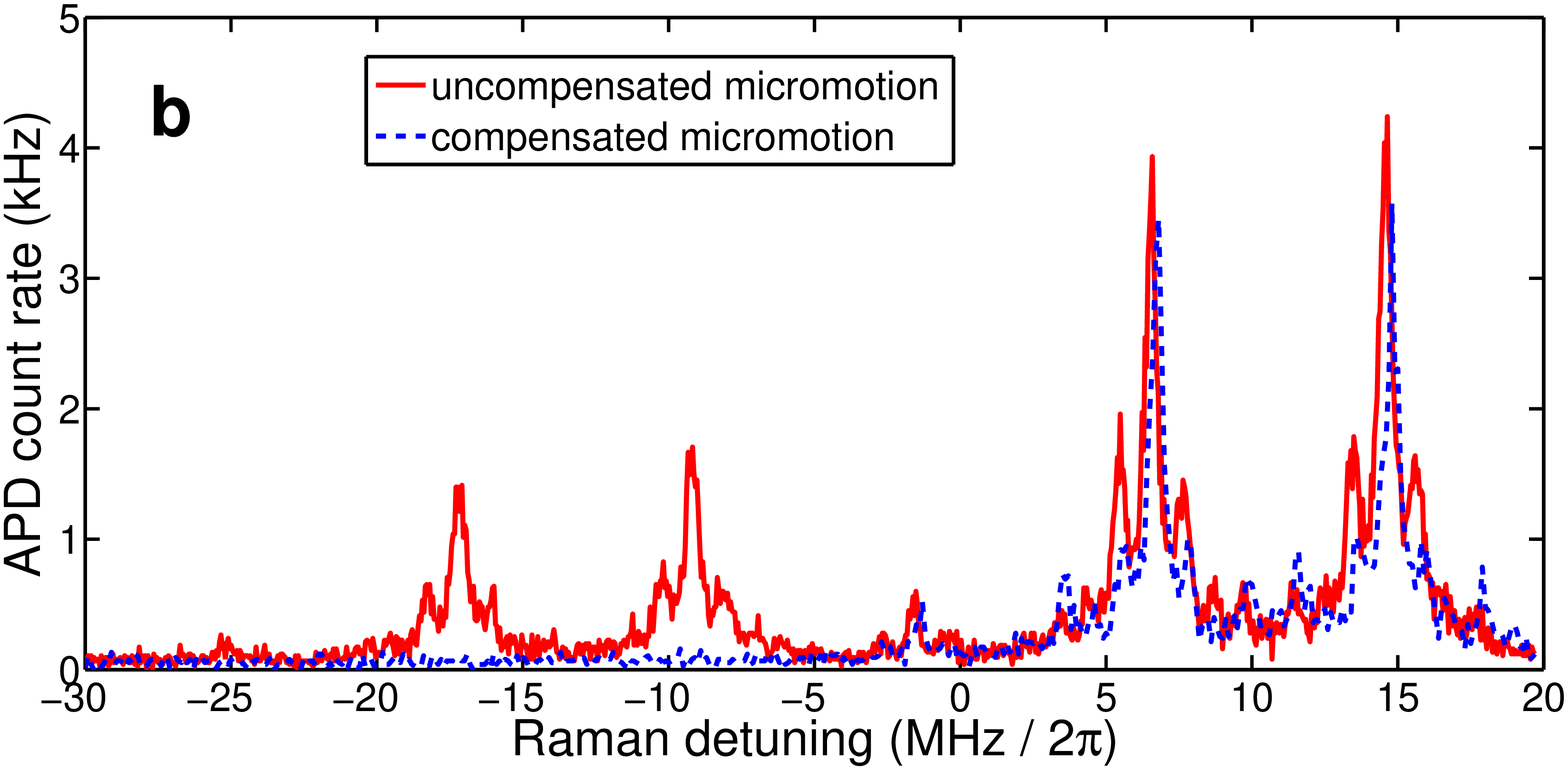}}
\caption{Motional sidebands resolved by Raman spectroscopy.  We drive cavity-assisted transitions using beam B and plot the APD count rate as a function of drive laser detuning. \textbf{(a)} We drive the \sohmmoh$\leftrightarrow$\xspace\pthmmth$\leftrightarrow$\xspace\dfhmmfh transition (also the rightmost transition of Fig.~\ref{fig_raman_spectrum_sigma}) and plot the count rate after Doppler cooling (red solid line) and axial sideband cooling (blue dashed line) of the ion.  The Rabi frequency  $\Omega_\mathrm{drv} = 2 \pi \times 33$~MHz.  Note the suppression of the red axial sideband and the reduction of the blue axial sideband after sideband cooling.  \textbf{(b)} Micromotion sidebands, offset from the primary spectrum by the trap drive radiofrequency of 23.4~MHz, are observable (red line) but can be suppressed (blue line) by applying DC voltages to compensation electrodes. For this data, $\Omega_\mathrm{drv} = 2 \pi \times 92$~MHz.}   
\label{fig_raman_spectrum_motion}     
\end{figure}
Motional sidebands are driven by beams A and B due to the nonzero projection of their $\vec{k}$-vector to all three motional axes.  
The axial and radial Lamb-Dicke parameters for both beams are given by 
$\eta_\mathrm{a} = 0.12$ and $\eta_\mathrm{r} = 0.05$. In Fig.~\ref{fig_raman_spectrum_motion}a, we plot the spectrum of beam B in a narrow scan of $\delta_\mathrm{drv}$ across the \sohmmoh$\leftrightarrow$\xspace\pthmmth$\leftrightarrow$\xspace\dfhmmfh transition.  
If the ion is sideband-cooled to the axial ground state before the drive laser is applied, we see that the axial red sideband is suppressed.
As the blue sideband height is proportional to $\sqrt{n_{\mathrm{a}}+1}\eta_{\mathrm{a}}$, where $n_{\mathrm{a}}$ is the ion's axial phonon number, the height of the blue sideband also decreases by a factor of $\frac{1}{\sqrt{n_{\mathrm{a}}+1}}$ \cite{Leibfried03}.

Moreover, we are able to drive micromotion sidebands at frequencies shifted by $\nu_{\mathrm{trap}} = 23.4$~MHz from those of the primary spectrum. Typically, micromotion is compensated in the experiment so as to suppress sidebands on the quadrupole transition, which also results in suppression on the cavity Raman transition.  For comparison, however, the Raman spectrum of beam B is shown in Fig.~\ref{fig_raman_spectrum_motion}b also for the case in which micromotion has not been properly compensated.

Evidence of motional sidebands is intriguing because of prospects for using the motional states of the ion to construct a quantum interface \cite{Parkins99a}. However, since the effective Raman coupling $\Omega_\mathrm{eff}$ depends on the Lamb-Dicke parameter $\eta$ to first order, but $\gamma_\mathrm{eff}$ depends on $\eta$ to second order, the ratio $\Omega_\mathrm{eff}  / \gamma_\mathrm{eff}$ is reduced by $\eta$ for motional sidebands.
For the intermediate-coupling parameters of our setup, this reduction means that atomic decoherence would play a significant role.

\subsection{Orthogonally polarized single photons}
\begin{figure}
\resizebox{0.48\textwidth}{!}{%
  \includegraphics[width=0.48\textwidth]{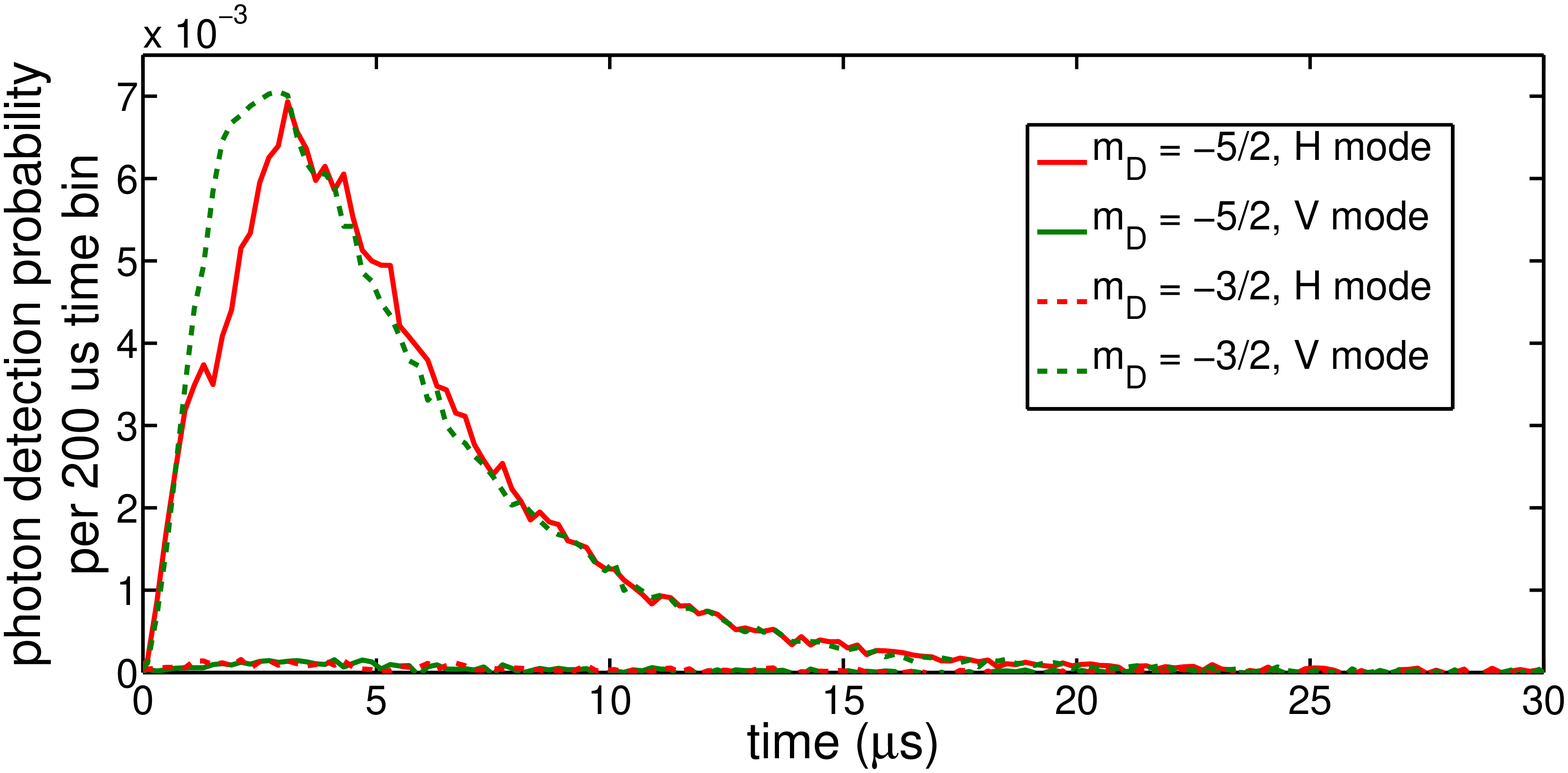}
}
\caption{Pulse shapes of single photons generated on two Raman transitions:  \sohmmoh$\leftrightarrow$\xspace\pthmmth$\leftrightarrow$\xspace\dfhmmfh (straight line) and \sohmmoh$\leftrightarrow$\xspace\pthmmth$\leftrightarrow$\xspace\dfhmmth (dashed line). Each Raman transition is driven in a separate experiment. The probability to detect a photon per 200~ns time bin is plotted as a function of time  after the drive laser is switched on.  Photons from the $H$ (red) and $V$ (green) cavity modes are detected separately, demonstrating the orthogonal polarizations of the two photon channels.}
\label{fig_photon_wavepacket}     
\end{figure}

Having identified our target transition frequencies in the spectra of Fig.~\ref{fig_raman_spectrum}, we now generate single photons in the cavity on demand at both frequencies.  We follow a procedure similar to that of Ref. \cite{Barros09}, in which photons were generated on a single \ssohtsdth transition.  After 600 \mus of Doppler cooling the ion, optical pumping for 150 \mus prepares the ion in the \sohmmoh state. A drive laser pulse of 80 \mus duration with Rabi frequency  $\Omega_\mathrm{drv} = 2 \pi \times 106$~MHz generates a single photon with an overall efficiency of 4.2\%, i.e., we record a detection event on one of the APDs in 4.2\% of all attempts. Subsequently, the repump lasers transfer all population back to the \ms manifold. In Fig.~\ref{fig_photon_wavepacket}, we plot the single-photon temporal pulse shapes for both transitions, driven one at a time.  Again, the polarization components of each photon pulse are separated in the detection path, which allows us to establish that the photon polarizations are orthogonal. In 1\% of cases, we observe photons with incorrect polarization, which we attribute to misalignment of the measurement axes with the cavity axes of polarization.

In order to avoid entanglement of the photon's polarization degree of freedom with its time-bin degree of freedom, it is necessary to overlap the two temporal pulse shapes.  Here, we can take advantage of the fact that the pulse shapes can be tuned by adjusting parameters of the drive laser, specifically, the Rabi frequency $\Omega_\mathrm{drv}$ and detuning $\delta_\mathrm{drv}$ \cite{Keller04}.  The pulse shapes of Fig.~\ref{fig_photon_wavepacket} already have substantial overlap; we expect to improve this by tuning the experiment parameters while addressing both transitions simultaneously.  

\subsection{Coherent manipulation of the atomic qubit}
\begin{figure}
\centering
\resizebox{0.5\textwidth}{!}{%
  \includegraphics[width=0.5\textwidth]{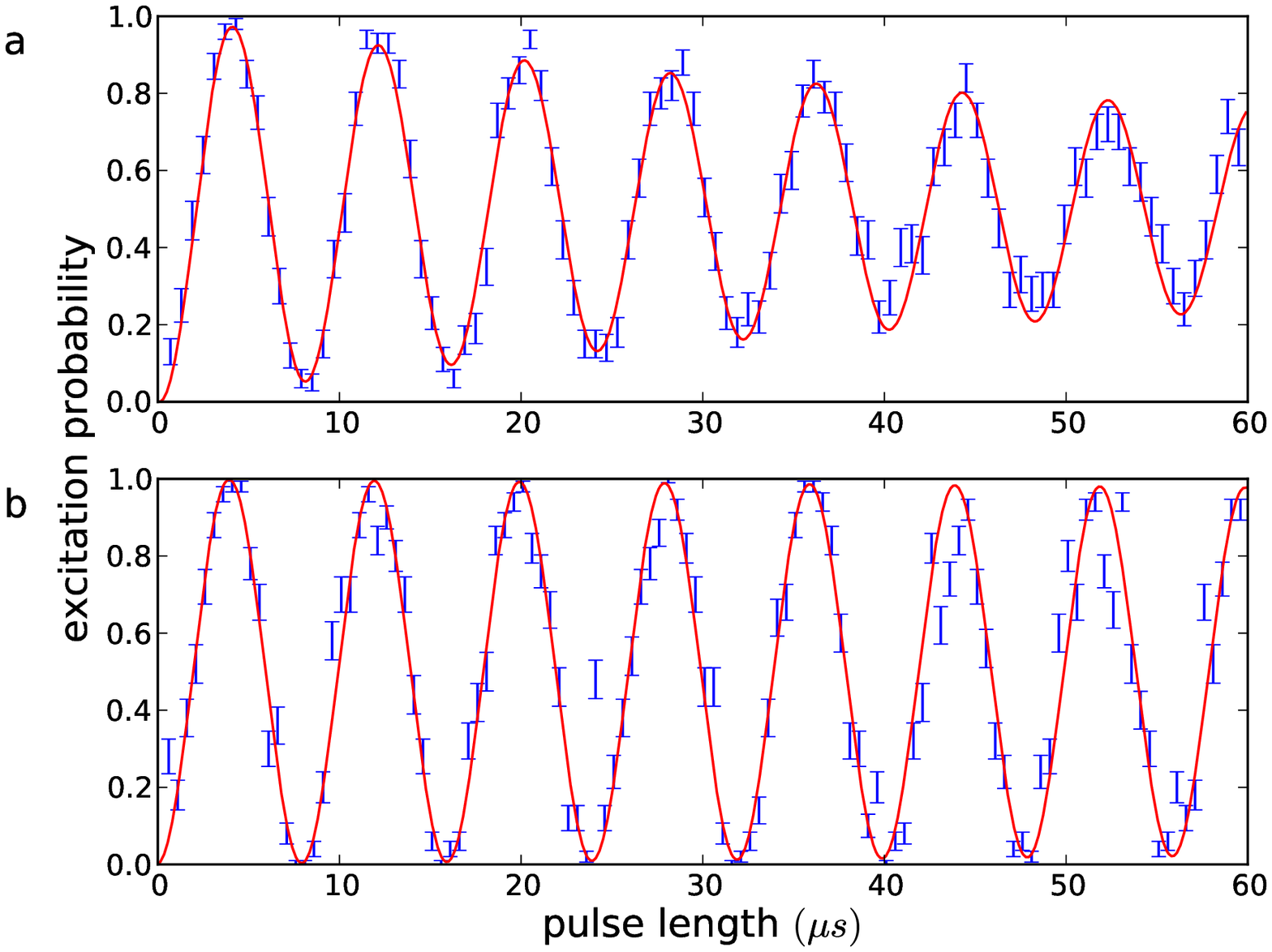}
}
  \caption{Rabi oscillations on the \ssohtsdfh quadrupole transition, driven at 729~nm. We plot the excitation probability of the \sdfh state as a function of pulse length. \textbf{(a)} After Doppler cooling, a range of phonon number states are occupied. Different frequencies result in a damped oscillation. \textbf{(b)} Following axial and radial sideband cooling, only one frequency component contibutes to the oscillation.}
\label{fig_rabi_flops}     
\end{figure}

An ion-photon interface requires coherent manipulation of the atomic qubit.  We have identified an atomic qubit based on two states in the \sdfh manifold,  \dfhmpmth and \dfhmpmfh.  As coherent operations are performed on the \ssohtsdfh quadrupole transition at 729~nm, we must couple the two qubit states via a third state in the \ssoh manifold.

In order to demonstrate the transfer of population between one \ssoh state and one \sdfh state, we drive Rabi oscillations between states \sohmpoh and \dfhmpfh, plotted in Fig.~\ref{fig_rabi_flops}. In Fig.~\ref{fig_rabi_flops}a, Doppler cooling is applied before the excitation of the quadrupole transition, while in Fig.~\ref{fig_rabi_flops}b, sideband cooling in all three dimensions is performed in addition to Doppler cooling.  In the first case, the oscillations are damped, as the Rabi frequency depends on the phonon number $n$ to second order in the Lamb-Dicke parameter, and states with a range of $n$ values (and thus a range of Rabi frequencies) are occupied after Doppler cooling.   
In contrast, after sideband cooling of all three vibrational modes ($\bar{n}_\mathrm{a} = 0.04 \pm 0.03$, $\bar{n}_\mathrm{r1} = 0.1 \pm 0.1$, $\bar{n}_\mathrm{r2} = 1.0 \pm 0.4$), the oscillations have only one frequency component.  

\begin{figure}
\resizebox{0.5\textwidth}{!}{%
  \includegraphics[width=0.5\textwidth]{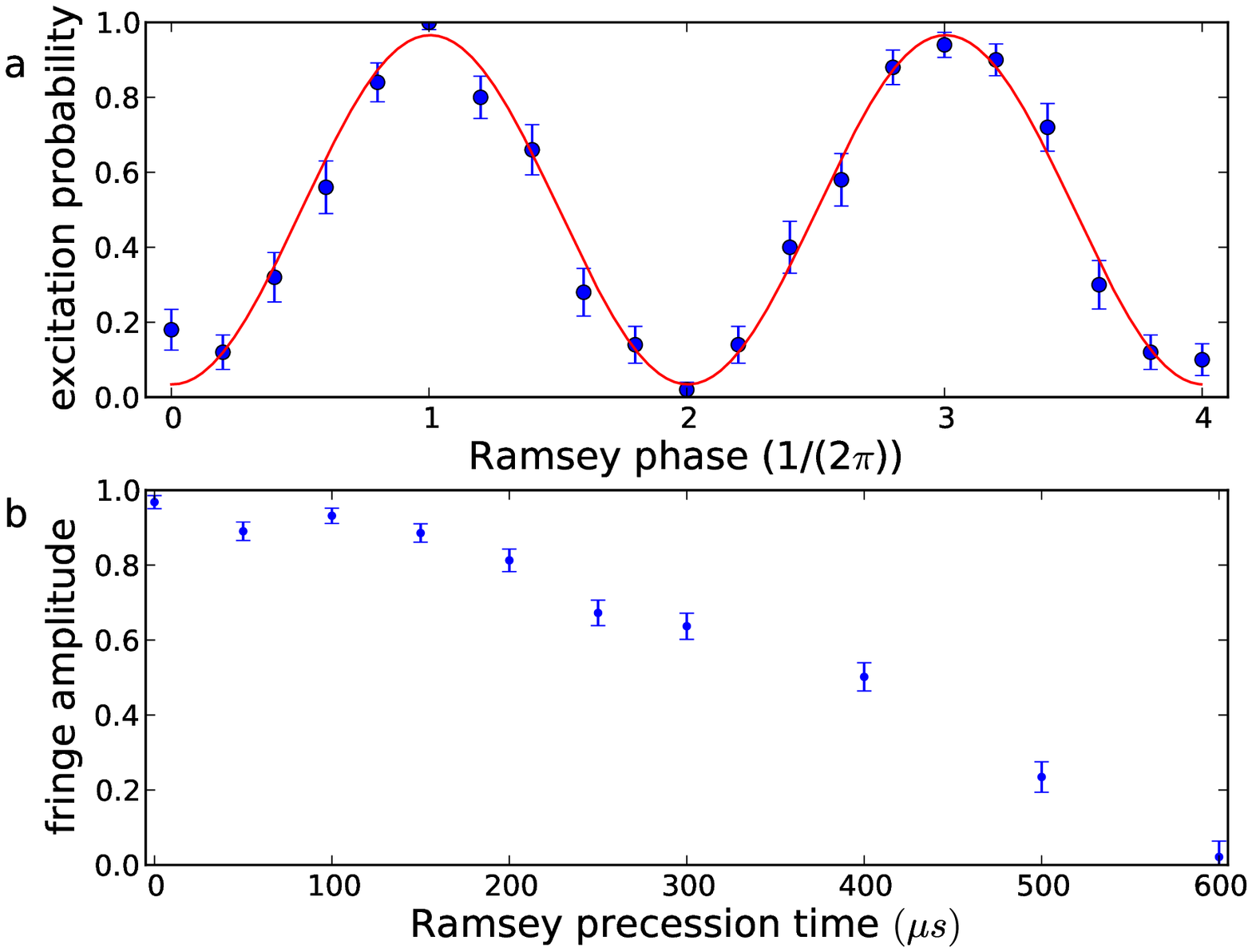}}
\caption{Coherence time of the atomic qubit. \textbf{(a)} Ramsey experiment for a superposition of \dfhmpfh and \dfhmpth, with precession time  $t_\mathrm{wait}$= 50 \mus.  The state is afterwards mapped to the \ssoh and \sdfh manifolds, and we plot the \sdfh excitation probability as a function of Ramsey phase, extracting an amplitude of 0.96. \textbf{(b)} Ramsey amplitude as a function of $t_\mathrm{wait}$. The coherence time is 250 \mus, limited by 50 Hz fluctuations of the magnetic field.}
\label{fig_ramsey_contrast}     
\end{figure}

It is important that we be able not only to manipulate the atomic qubit state but also to store information within it for extended times.  We investigate the atomic coherence time 
for a superposition of \dfhmpth and \dfhmpfh in the following Ramsey experiment.  After Doppler-cooling the ion and optically pumping it to the \sohmpoh state, a first $\pi/2$ pulse on the \sohmpoh\fm{\leftrightarrow}\xspace\dfhmpfh transition generates an equal superposition of \sohmpoh and \dfhmpfh. Next, a $\pi$ pulse on the \sohmpoh\fm{\leftrightarrow}\xspace\dfhmpth transition transfers the remaining \sohmpoh population entirely to \dfhmpth. After a precession time $t_\mathrm{wait}$, the same pulse sequence is applied in the reverse order, and the phase of the second $\pi/2$ pulse is scanned. In Fig.~\ref{fig_ramsey_contrast}a, we plot the Ramsey fringe for $t_\mathrm{wait}$= 50 \mus as an example. 

For each value of $t_\mathrm{wait}$, we observe a Ramsey fringe as a function of phase, from which we can extract the fringe amplitude. 
From the dependence of this amplitude on $t_\mathrm{wait}$, which we plot in Fig.~\ref{fig_ramsey_contrast}b, we infer a coherence time of 250 \mus, limited by slow fluctuations (50 Hz) of the magnetic field in the laboratory.  Since photon generation takes place in $\leq 20~\mu$s (Fig.~\ref{fig_photon_wavepacket}) and individual gate operations last a few $\mu$s (Fig.~\ref{fig_rabi_flops}), it is possible within this coherence time both to generate individual photons (or photon sequences) and to carry out multiple rotations of the atomic qubit.

\section{Proposed quantum interface implementations} \label{proposal_schemes}
  We conclude by considering possible realizations of a cavity-based ion-photon interface using $^{40}$Ca$^{+}$.

\subsection{Atom-photon entanglement}

Entanglement of single photons and single atoms has been previously demonstrated \cite{Blinov04,Volz06} and, in the case of ions, was subsequently the basis for entanglement of remote qubits conditioned on joint detection of photons \cite{Moehring07}.  In this context, coupling the ion to an optical cavity offers significant improvements in efficiency, as entangled photons are emitted with near-unit efficiency into the cavity mode, where they are easily collected.  Within a cavity, neutral atoms have mediated the entanglement between photons emitted in sequence \cite{Wilk07b}, but the use of an ion enables coherent manipulation and direct readout of the atomic state.

In our setup, we plan to initialize the ion in the state \sohmmoh.  A bichromatic drive pulse with $\sigma^{-}$-polarization will simultaneously excite the two Raman transitions identified in Section \ref{raman_transitions_in_ca}, generating the joint ion-photon state
\begin{align}
\frac{1}{\sqrt{2}} \left(\dfhmmfh \ket{H} + e^{i\phi} \dfhmmth \ket{V}\right), 
\end{align}
where $H$ and $V$ represent photon polarization states, $\phi$ is determined by the relative phase of the two drive frequencies, and we assume that the temporal shapes of Fig.~\ref{fig_photon_wavepacket} have been tailored to match one another. Photons generated along both paths have the same energy, determined by the cavity frequency.  

In principle, one can also generate an arbitrary entangled state through the selection of appropriate Rabi-frequencies $\Omega_\mathrm{drv}^{i=1,2}$ and detunings $\delta_\mathrm{drv}^{i=1,2}$.

Concerning laboratory implementation, it is important to note that the bichromatic pulse frequencies must be phase stable with respect to one another.  A second technical requirement is that optical elements in the cavity output path with polarization-dependent losses, such as dichroic elements for cavity stabilization, must only be integrated after the polarization analysis setup.

\subsection{Atom-photon state mapping}

Atom-photon entanglement can be generated using spontaneously emitted photons, as in Ref. \cite{Blinov04}.  However, an essential property of an atom-photon interface is the ability to map an arbitrary quantum state of the atom onto a photonic state \cite{Cirac97}, and here a coherent rather than a spontaneous process is required.  

To demonstrate such a mapping, we intend to initialize an arbitrary atomic state 
\begin{align}
\cos(\alpha) \sohmmoh + e^{i \phi} \sin(\alpha)  \sohmpoh
\end{align}
by first optically pumping the ion to a single \ssoh state, then coherently driving any pair of \ssohtsdfh transitions that couple the two ground states. Again, we will simultaneously excite two Raman transitions via bichromatic drive pulses; in this case, however, the transitions will have different initial states but will share a final state
\begin{align}
(\cos(\alpha) \ket{V} + e^{i\phi} \sin(\alpha) \ket{H})\dfhmmth
\end{align}
in which the atomic superposition has been transferred to the photonic qubit. 

One powerful application of this mapping is the sequential generation of photonic cluster states, recently proposed for quantum dot systems \cite{Lindner09}. In the context of our ion-cavity system, the atomic qubit acts as a memory, emitting a linear string of entangled photons, one by one.  Pairwise entanglement of photons can be generated by coherent manipulation of the ion between pulses.  Trapping multiple ions within the cavity mode enables the generation of higher-dimensional cluster states \cite{Economou10}. 

These implementations are ambitious but also feasible within the current experimental system.   The toolbox that we have outlined --- including the identification of Raman transitions between individual Zeeman states, the generation of orthogonally polarized cavity photons, and the manipulation of atomic qubit states --- provides the fundamental components for an ion-cavity quantum interface.

\section{Acknowledgments} \label{acknowledgments}
We thank K. Hammerer for helpful discussions.  We gratefully acknowledge support from the Austrian Science Fund (FWF), the European Commission (AQUTE),
and the Institut f\"ur Quanteninformation GmbH.  T.E.N. acknowledges support from a Marie Curie International Incoming Fellowship within the 7th European Community Framework Programme, and B.B. acknowledges support from the DOC-fFORTE fellowship of the Austrian Academy of Sciences.

\bibliographystyle{applphysb}
\bibliography{references/bibsonomy_cqed}



\end{document}